# Long-lived electron spins in a modulation doped (100) GaAs quantum well

J. S. Colton, D. Meyer, K. Clark, D. Craft, J. Cutler, T. Park, P. White

*Department of Physics and Astronomy, Brigham Young University, Provo UT*

We have measured $T_1$ spin lifetimes of a 14 nm modulation-doped (100) GaAs quantum well using a time-resolved pump-probe Kerr rotation technique. The quantum well was selected by tuning the wavelength of the probe laser. $T_1$ lifetimes in excess of 1 microsecond were measured at 1.5 K and 5.5 T, exceeding the typical $T_2^*$ lifetimes that have been measured in GaAs and II-VI quantum wells by orders of magnitude. We observed effects from nuclear polarization, which were largely removable by simultaneous nuclear magnetic resonance, along with two distinct lifetimes under some conditions that likely result from probing two differently-localized subsets of electrons.

## I. INTRODUCTION

Since the initial proposal of spin-based quantum computing[1] and the discovery of very long inhomogeneous dephasing spin lifetimes ($T_2^*$) in GaAs,[2] a tremendous amount of research effort has been put forth to better understand the interaction of electronic spin states ("spins") with each other and with their environment, and to create structures on the nanoscale that allow for better control and study of the spins.[3] Among the key requirements for semiconductor spintronic devices is an understanding of the spin dephasing mechanisms in semiconductors.[4] Optical techniques for interacting with spins in semiconductor heterostructures are powerful tools for the initialization, manipulation, and study of spin dynamics.[5] GaAs/AlGaAs heterostructures are ideally suited for such experiments, as GaAs is a direct-gap semiconductor with well-known selection rules connecting optical polarization to the spin degree of freedom. Additionally, the band-gaps of GaAs/AlGaAs heterostructures readily match commercially-available lasers such as Ti:sapphire, which allows for resonant excitation and detection of the electronic spins.

Many experimental studies on GaAs have focused on lightly doped *n*-type bulk material, where electrons localize on donor sites at low temperature. Spin lifetimes much longer than the optical lifetimes can be obtained with these doped electrons. A wide variety of experimental techniques have been employed to study this type of bulk material, including (but not limited to) Hanle effect depolarization,[6] time-resolved Faraday or Kerr rotation,[2,7] optically-detected electron spin resonance,[8,9] time-resolved decay of photoluminescence polarization[10,11] or polarization-dependent luminescence,[12] optically-controlled spin echo,[13] Kerr rotation imaging,[14,15] and spin noise spectroscopy.[16]

Other experimental studies have focused on InAs or InGaAs quantum dots embedded in a GaAs barrier, again with doped electrons added to the dots to allow the electron spin information to be preserved beyond the radiative recombination time. In self-assembled quantum dots, for example, optical techniques have allowed the electron spins to be precisely controlled on time scales of micro- or milliseconds.[17-20]

Bridging the gap between bulk material and 0D quantum dots, 2D systems can serve as well-defined model systems for studies in spin dynamics. Early studies of spins in quantum wells often focused on exciton dynamics.[21] However, time-resolved studies have also served to shed light on properties of the electrons in GaAs quantum wells, allowing the dephasing of spins in subnanosecond[22] and nanosecond[23,24] time scales to be directly measured. The longest spin dephasing times in GaAs quantum wells have ranged from 10-30 ns.[24,25] Other promising results have been obtained in II-VI quantum wells, where spin dephasing times of 30 ns have also been observed through various techniques[26,27] and some degree of optical control of spins has been established.[28]

Throughout these previous experiments, the spin lifetimes in quantum wells that have been the focus of research have nearly always been the $T_2^*$ lifetimes, also called the inhomogeneous dephasing times. By contrast, in this paper we present experimental measurements of $T_1$ spin lifetimes, also known as spin flip times. While $T_2^*$ is measured with the field perpendicular to the spin orientation, $T_1$ is measured with a parallel field. $T_2^*$ and $T_1$ are generally considered lower and upper bounds for $T_2$, the true dephasing time.

In this work we have measured the $T_1$ spin lifetime of a 14 nm GaAs quantum well using a time-resolved pump-probe Kerr rotation technique. The spin lifetimes were quite long—tens and hundreds of nanoseconds at most fields (from 0-7 T) and temperatures (1.5 and 5 K), and exceeding one microsecond at the lowest temperature and highest field. This paper is structured as follows: Sec. II describes the sample.

Sec. III discusses the polarization and detection scheme, along with some wavelength-dependent results. Sec. IV gives details on our experimental setup for spin lifetime measurements. The main experimental results and discussion are found in Sec. V, after which we provide some discussion in Sec. VI. We conclude in Sec. VII.

## II. SAMPLE

We studied a 14 nm wide GaAs quantum well which was grown through molecular beam epitaxy and modulation doped with silicon donors to produce a carrier concentration of $n = 3 \times 10^{10}$ cm$^{-2}$ in the well. It is part of a multi-quantum well sample containing five total wells with thicknesses of 2.8, 4.2, 6.2, 8.4, and 14 nm. More details on the sample's structure and electronic properties can be found in Ref. 29. The 14 nm well was selected by tuning our laser to the optical transition of that well, approximately 807 nm. As with experiments in other $n$-type bulk, quantum dot, and quantum well samples mentioned in the Introduction, the doping allows spin information to be preserved through the ground state electrons.

This particular well of this particular sample has been the study of other spin-related investigations by our group and others, including Hanle effect measurements of $T_2^*$,[30] time resolved Kerr rotation measurements to study optical initialization and $T_2^*$ lifetimes,[23] and optically-detected electron spin resonance measurements which manipulated spin states with microwaves.[31]

## III. SPIN POLARIZATION AND DETECTION

The modulation doping causes a background of electrons to exist in the well, which can interact with optically-injected excitons to form trion states. We consider only the lowest energy, singlet trions, where two electrons of opposite spin form a bound state with a hole which can be either spin-up or spin-down. The hole spin can be either $\pm 3/2$ or $\pm 1/2$, depending on whether it is a heavy or light hole. Because the two electrons in the singlet state have opposite spins, the overall spin of the trion follows the hole spin and is either $\pm 3/2$ or $\pm 1/2$. The details of trion formation rely critically on whether the optical photon has spin +1 (labeled $\sigma^+$) or spin −1 (labeled $\sigma^-$), and are depicted in Fig. 1.

The polarization of the ground state electron spins in doped quantum wells has typically been done through resonant excitation of a trion state.[23,32,33] That mechanism relies on fast hole relaxation in the excited state: with $\sigma^+$ photons (for example) resonant with the heavy hole trion transition, electrons are taken out of the +1/2 state into the +3/2 heavy hole trion. The rapid hole relaxation causes the trion population to be equalized between the +3/2 and −3/2 states. In GaAs this can occur extremely state, and a ground state spin polarization occurs. (If there were no hole spin flips, the +3/2 trion would simply decay back into the +1/2 ground state and no ground state spin polarization would accumulate.)

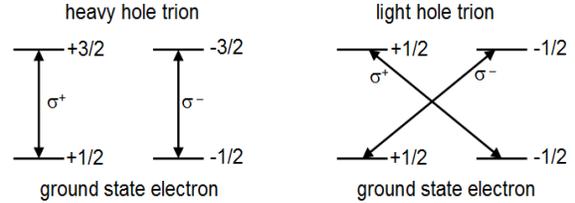

FIG. 1. Heavy and light hole trion transitions and selection rules. The trion forms when a ground state electron (+1/2 or −1/2) combines with an optically-injected electron hole pair. Because the two electrons are in a singlet state, the spin state of the trion matches the spin state of the hole (+3/2 or −3/2 for the heavy hole trion; +1/2 or −1/2 for the light hole trion). Photon spin states of +1 and −1 are indicated by $\sigma^+$ and $\sigma^-$ respectively.

Our approach was slightly different. We performed a two-color experiment with pump and probe photons having different energies. Although our probe laser was resonant with a trion transition (details below), our pump laser (781 nm) was at a much higher energy. Our pump laser therefore excited both heavy and light hole trions simultaneously. Again considering $\sigma^+$ photons: they will excite heavy hole trions and pump spins out of the +1/2 ground state as described in the previous paragraph; however, they will also pump spins out of the −1/2 ground state by exciting light hole trions. As in the case of $n$-type bulk material—where the heavy and light hole states are degenerate at the band edge and are thus always excited simultaneously with a pump laser—we rely on unequal transition probabilities for the heavy hole state compared to the light hole state to generate a net spin polarization for the ground state electrons. Two-color experiments have been done in II-VI quantum wells in order to separate the effects of detecting the exciton vs. trion transitions,[27,33,34] and have been proposed for use in a non-resonant pumping scheme such as we employed,[34] but we are not familiar with any other actual two-color experiments in GaAs quantum wells.

To detect the persisting electron spin polarization we tune the probe laser to be resonant with the trion transitions. This is quite similar to those groups cited above who employed a single-color resonant pump-probe scheme. We use the Kerr effect, i.e. the rotation of the angle of polarization of our linearly polarized probe beam, to detect the ground state spin population. Under typical conditions when the probe laser is tuned resonant with the quantum well's optical transition there is a clear signal with two features; see Fig. 2. The feature at 807.3 nm is from the light hole trion; the one at 808.1 nm is from the heavy hole trion. (As noted by Kennedy et al., the heavy hole *exciton* transition likely partially overlaps the light hole trion.[23]) The peaks are opposite in sign because of the opposite selection rules depicted in Fig. 1.

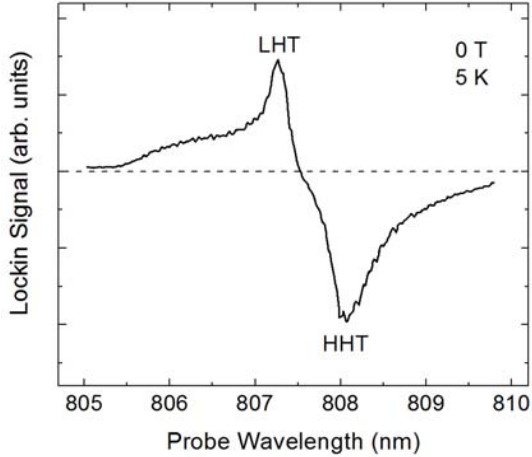

FIG. 2. Kerr rotation signal taken at 0 T, 5 K, as a function of probe laser wavelength. Data was obtained with pump and probe pulses each set for 50% duty cycle and overlapping each other in time. The dashed line indicates the zero position; the two peaks have opposite sign due to the optical selection rules. LHT and HHT label the light and heavy hole trion peaks, respectively.

In principle, the ground state electron spin polarization should be able to be measured through either the heavy hole or light hole trion transitions; in practice it proved easier for us to set our probe laser to the light hole trion transition because at some fields the heavy hole trion feature was difficult to observe; see Fig. 3a for a collection of wavelength-dependent data at various fields. The peak positions of Fig. 3a are summarized in Fig. 3b. The peak positions of the heavy and light hole trions as a function of magnetic field follow the well-known quadratic "diamagnetic shift," in this case given by the following equations fitted from the data with energies in eV and fields in T:

$$E_{LHT} = 1.5360 + 4.289 \times 10^{-5} B^2 \quad (1a)$$
$$E_{HHT} = 1.5344 + 4.289 \times 10^{-5} B^2 \quad (1b)$$

The two trion peaks maintain a constant separation of 1.57 meV which is in good agreement for the LHT-HHT separation reported in Ref. 23.

## IV. EXPERIMENTAL SETUP

To study the $T_1$ behavior of the ground state electron spins, we used a two-color pump-probe technique described in detail in Ref. 7. The magnetic field is oriented in Faraday (longitudinal) geometry, with the field parallel to the spin alignment. As mentioned in Section III, the spins are aligned using a circularly polarized pump laser and detected via the Kerr rotation of a linearly polarized probe laser. Both pump and probe lasers are pulsed, and the delay between the two of them is varied. This is similar to the traditional time resolved

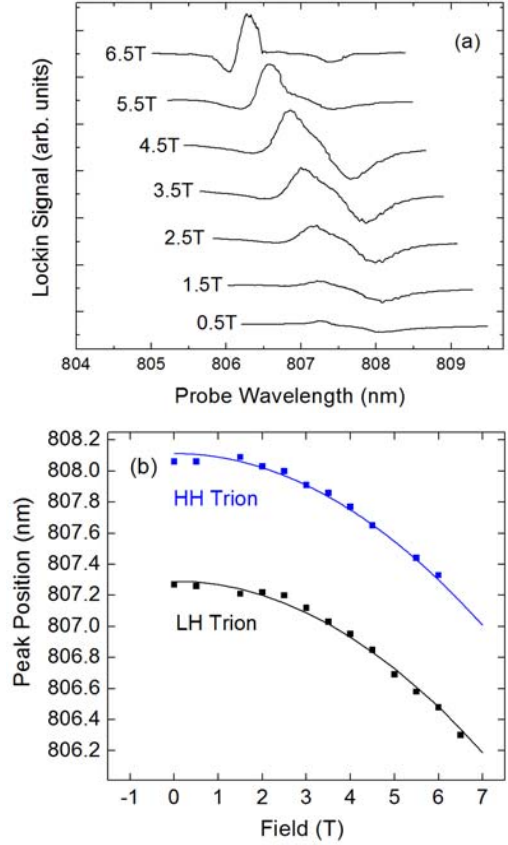

FIG. 3. (a) cw Kerr rotation signal vs. probe laser wavelength at 5 K for selected fields. Data was obtained under the same conditions as Fig. 2. (b) Summary of peak positions, fitted to a quadratic function in energy.

Kerr (or Faraday) rotation (TRFR) technique used by many to measure the inhomogeneous dephasing lifetime, $T_2^*$, of various semiconductors. However, in order to access the much longer lifetimes that are involved with $T_1$ as opposed to $T_2^*$, we employed electronic gating of pump and probe pulses instead of a mechanical delay line to vary the delay. Also, because the spins are parallel to the external field, we do not see the precession oscillations that are a hallmark of the traditional TRFR technique.

The pulses in the probe beam, a tunable cw Ti:sapphire laser, were produced with an acousto-optic modulator (AOM). Because the probe beam was quasi-cw—only pulsing on the time scales of tens of nanoseconds in response to our AOM—its bandwidth is essentially infinitely narrow on the scales of Figs. 2 and 3 and excellent wavelength resolution was achieved. The pump beam, a fast diode laser, was modulated on/off via a direct modulation input. The two beams were synchronously controlled with a two-channel pulse generator. To separate out the spin effects from sources of noise and to reduce dynamic nuclear polarization, we modulated the helicity of the pump laser from $\sigma^+$ to $\sigma^-$ with a 42 kHz photo-elastic modulator and detected the signal with a lockin amplifier referenced to that frequency. The lockin signal is proportional to the spin polarization of the electrons in the sample.

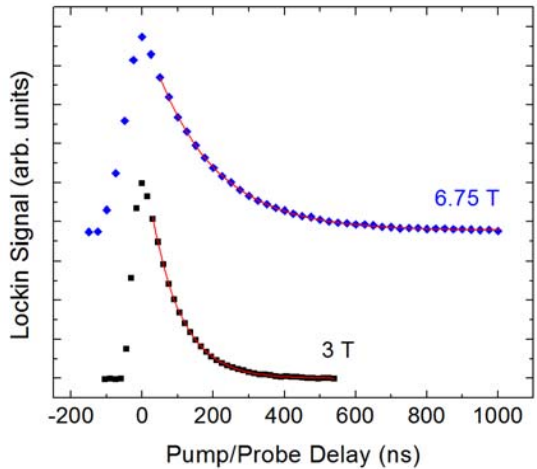

FIG. 4. Representative 5 K data taken at 3 T and 6.75 T: spin polarization vs. delay between pump and probe pulses. The raw data for the spin decays (points) was fitted to exponential decays (solid curves), yielding spin lifetimes of 84.5 ns and 169 ns for the 3 T and 6.75 T data, respectively. The 6.75 T data has been shifted vertically for clarity.

The pump beam was set to 25 mW unpulsed and was focused (partially) to a diameter of 0.22 mm. The probe beam was set to a diameter of 0.21 mm and its power was either 3.5 mW unpulsed (for the 5 K data) or 2 mW unpulsed (for the 1.5 K data). The overall time for a pulse repetition cycle was approximately six times the decay time, and pulse widths were set to give the pump a duty cycle of 4% and the probe a duty cycle of 2%.

The sample was placed in a superconducting electromagnet with integrated cryostat where fields up to 7 T and temperatures down to 1.5 K could be investigated.

## V. RESULTS

For a given set of experimental parameters, the delay between pump and probe was varied in order to trace out the decay of spin polarization. As the delay is varied, the probe pulse temporally "enters" the pump pulse, causing the lock-in signal to rapidly rise, then "exits" the pump pulse causing a decrease in signal. Any signal which exists after the probe pulse has exited the pump pulse is a result of persisting spin information. The polarization typically decays exponentially as:

$$P = P_0 \exp(-t/T_1) \qquad (2)$$

See Fig. 4 for two representative decays and their fits, which in this case yielded spin lifetimes of 84.5 ns and 169 ns for the 3 T and 6.75 T data, respectively.

Figure 5 displays a summary of our spin lifetime results for spin decays measured at both 5 K and 1.5 K.

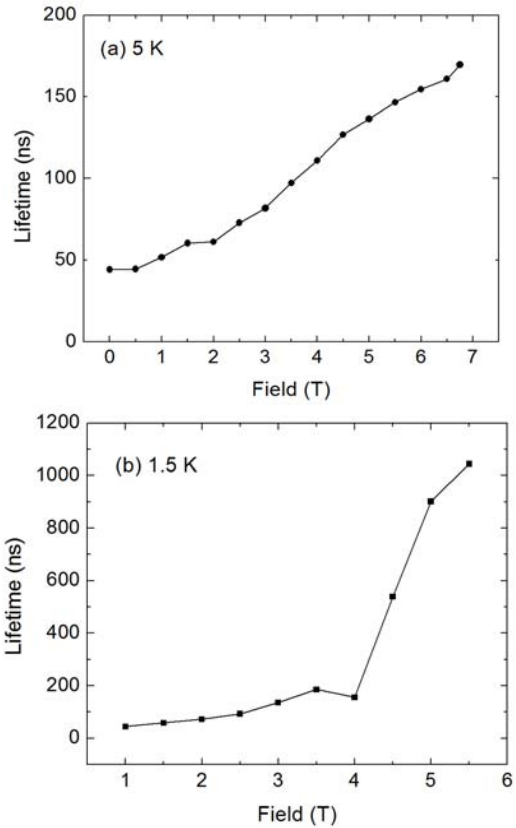

FIG. 5. Measured electron spin lifetimes as a function of magnetic field for (a) 5 K and (b) 1.5 K. For the 1.5 K data, the scans for fields at 4 T and above were done with rf applied to remove nuclear polarization (which may have caused some heating of the sample).

Spin lifetimes from 44 ns to 170 ns were measured at 5 K, and lifetimes from 44 ns to 1040 ns were measured at 1.5 K. These lifetimes far exceed the $T_2^*$ value of 2.5 ns reported by Kennedy et al. for this particular quantum well (at 0 T, 6 K),[23] and are also much longer than the longest lifetimes (also $T_2^*$) of ~30 ns reported for *any* quantum wells of which we are aware, as referenced in the Introduction. $T_1$ is generally considered an upper bound for the true coherence time $T_2$, and in (100) quantum wells $T_2$ is expected to be on the same order of magnitude as $T_1$.[41] Therefore these long $T_1$ results may be an indication that spin coherence can persist in quantum wells much longer than has generally been considered to be the case, and that e.g. spin echo experiments should be pursued in quantum well samples.

Nearly all of the raw data followed precise exponential decays like the two representative plots in Fig. 4. However, at 1.5 K there were some field points which did not follow a simple exponential decay. For those points, the lifetime that is plotted is simply the $1/e$ fall-off point for the raw data after the peak.

One reason for non-exponential decays is the presence of a nuclear spin polarization. Nuclear polarization is expected to arise whenever the electron spin polarization is far from thermal equilibrium. This is especially the case for us for the high field, low temperature situations. From simple Boltzmann statistics, the polarization of a two level spin system is:

$$P = \tanh(g\mu_B B / 2k_B T) \quad (3)$$

The *g*-factor for this well was obtained in previous spin resonance experiments,[31] |*g*| = 0.346, so at 1.5 K the thermal equilibrium polarization of the electrons will be 30%, 37%, and 43%, for fields of 4, 5, and 6 T, respectively. However, the pump laser—with its helicity modulated between $\sigma^+$ and $\sigma^-$ as described above—will be driving the electron polarization towards 0%, at least on time scales long compared to the modulation time of (42 kHz)$^{-1}$. As the electrons are driven toward zero polarization, they will attempt to return to their thermal equilibrium value by interacting with the nuclear spin bath via the hyperfine interaction. This will polarize the nuclear spins to some degree. Polarized nuclei impact the electrons via the Overhauser effect and generate an effective field for the electrons. This effective field can vary both physically across our laser beams, as well as temporally during our scans, and can change the measured spin response in unpredictable ways.

That there *is* substantial nuclear spin polarization present in the material under some conditions was evident. Fig. 6 displays two spin decays taken at 5.5 T and 1.5 K, under nearly identical conditions. The only difference is that the solid curve was performed while rf was applied to a Helmholtz coil surrounding the sample, sweeping through the frequencies needed for nuclear magnetic resonance (NMR) of the four nuclear isotopes present in the quantum well and barrier: $^{75}$As, $^{69}$Ga, $^{71}$Ga, and $^{27}$Al. This was done via a function generator with customizable frequency modulation. With rf applied to remove built-up nuclear polarization, a relatively normal decay was observed. However, without rf the shape was both non-exponential and non-reproducible. As can be seen, for the data presented in Fig. 6 the spin polarization initially remained *constant* as the probe pulse begins to arrive after the pump. Something is changing inside the sample (i.e., the nuclear spins) in order to preserve the electron spin polarization! This is very reminiscent of the "spin dragging" effect that has been observed in electron spin resonance of bulk GaAs[35] and GaAs-based quantum dots,[36] where nuclear polarization has also been seen to adjust to keep the electronic polarization constant. These nuclear polarization effects were seen for all of the 1.5 K data at fields of 4 T and higher. The lifetimes plotted in Fig. 5b for these fields are for the "rf on" set of measurements. Although the decays for these points looked reasonable, as in the rf on curve of Fig. 5b, they could not be fitted to simple exponential decays—indicating that our removal of the effects from nuclear polarization was incomplete. The rf likely also caused some small heating of the sample, which could explain

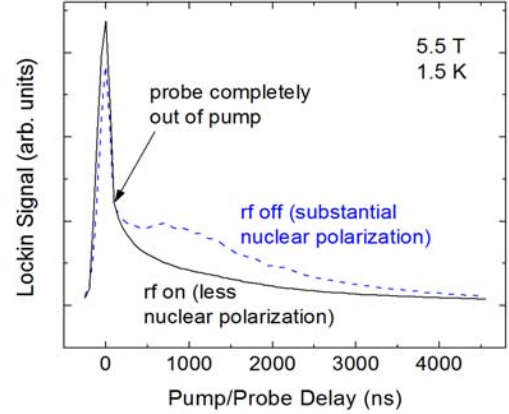

FIG. 6. Spin decays measured at 5.5 T and 1.5 K. The solid and dashed lines are for conditions with and without rf applied to depolarize the nuclear spins.

the unexpected decrease in lifetime in Fig. 5b going from 3.5 T (no rf) to 4 T (with rf).

## VI. DISCUSSION

To discuss our spin relaxation results further, we first review some of the theoretical work on spin lifetimes in quantum wells. Spin scattering in quantum wells was first discussed by D'yakonov and Kachorovskii (DK).[37] In GaAs-based quantum wells, the lack of bulk inversion symmetry leads to spin-splitting of the conduction band. This spin splitting can be regarded as an internal magnetic field, about which electrons precess between momentum scattering events. This leads to information loss about the initial spin state and is called the D'yakonov-Perel mechanism. D'yakonov and Kachorovskii analyzed that mechanism in the context of quantum wells to obtain this result for the spin lifetime:

$$\tau_s = \frac{E_g \hbar^2}{\alpha^2 E_1^2 k_B T} \frac{1}{\tau_v}, \quad (4)$$

where $E_g$ = is the band gap energy, $E_1$ is the electron's quantized energy in the well, $T$ is the temperature, $\tau_v$ is the momentum scattering time (which also depends on temperature), and $\alpha$ is a parameter related to the spin splitting of the conduction band. An important result is that generally speaking a short momentum scattering time ($\tau_v$) will result in a long spin lifetime ($\tau_s$), and vice versa. In asymmetric quantum wells, there is additionally a structural inversion asymmetry, which can add to or subtract from the effects of the bulk inversion asymmetry. This has recently been used in an experiment by Balocchi et al. to partially cancel the relaxation term from bulk inversion asymmetry (Dresselhaus) with the term from structural inversion asymmetry (Rashba).[24]

The general theoretical approach is therefore often to model the momentum-scattering mechanisms that contribute to $\tau_v$; for example, Bastard and Ferreira used the DK theory to describe ionized impurity scattering, often the most efficient scatterers at low T.[38] They found that $\tau_\nu$ shortens considerably at low temperatures due to inefficient screening, yielding spin flip times that are the longest for wide wells and low temperatures. For their particular impurity concentration and screening model, they predicted $\tau_s$ to be 2.5 ns at 10 K for a 15 nm GaAs/AlGaAs well and there data points suggest that $\tau_s$ should increase rapidly with a decreasing temperature. A simple extrapolation of their data suggests a factor of 10 or 100 increase in lifetime as temperature decreases to 1.5 K. Bastard extended the DK theory to a high magnetic field situation using Landau levels and a point-like defect model for the scatterers, to obtain a prediction of 1-2 ns for a 9 nm well for fields between 6 and 15 T and a $B^{1/2}$ dependence of lifetime on field.[39]

Experimentally, Terauchi et al. measured spin lifetimes at 0 T and 300 K in a series of 7.5 nm multi-quantum well samples, and verified the $\tau_s \sim 1/\tau_\nu$ prediction of the DK theory, although the spin lifetimes were about an order of magnitude higher than the theory predicted.[40] Lau et al. built upon the DK theory in two papers, using a 14 band **k·p** model to describe bulk[41] and structural inversion asymmetry,[42] and overcame the order of magnitude discrepancy that had been seen. $T_1$ and $T_2$ were predicted to be the same order of magnitude, with $T_2$ ranging from $2T_1/3$ to $2T_1$ in (100) wells depending on the value of $\alpha$. Their calculated $T_1$ values matched the room temperature experiments of Terauchi et al.,[40] and the $T_2$ values matched the original experiments of Kikkawa and Awschalom[2] for temperatures of 100 K and above. They issued the disclaimer, however, that their theory might not be applicable at lower temperatures.

More recent theory on spin relaxation in $n$-doped quantum wells is sparse, the work of Harmon et al. being a notable exception.[43] Their work focuses on spin dephasing from the hyperfine interaction, applicable to $T_2^*$ but not to $T_1$. They also explicitly account for doping via donors inside the well, and mention that their theory is consequently not applicable for modulation doped wells (such as used in our experiment).

Considering our measured value of 44 ns for $T_1$ at 0 T and 5 K, our results seem fairly solidly in the Bastard and Ferriera regime (if the low temperature extrapolation of their data is to be believed), and likely indicates that ionized impurity scattering within the DK model is our primary relaxation mechanism. Our quantum well is modulation doped, so there are no *intentional* impurities in the well, but this sample did have a slight $n$-type background. Our spin lifetime increase with magnetic field did not exactly follow the $B^{1/2}$ prediction of Bastard, but our lifetimes did increase nearly monotonically with field as Bastard's theory predicts. Comparing our results to other experimental results, one would expect our 0 T, 5 K value for $T_1$ to match fairly closely the 0 T, 6 K value for $T_2^*$ of Kennedy et al.,[23] since the sample is the same and $T_2^*$ and $T_1$ are equivalent in the absence of a magnetic field. However, the value of Kennedy et al. seems to have been obtained from a fit of a decay measured only between 0 and 1.5 ns, and consequently their value of 2.5 ns may not be completely trustworthy.

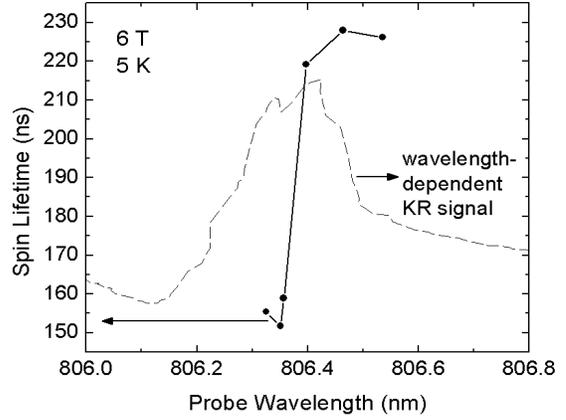

FIG. 7. Spin lifetimes as a function of probe laser wavelength, for 6 T and 5 K. Note the abrupt shift in lifetime over a very small wavelength range. For reference, the wavelength-dependent Kerr rotation for these conditions, showing the light hole trion peak, is also displayed (dashed).

Finally, in two previous papers on $T_2^*$ lifetimes in II-VI quantum wells, two distinct spin lifetimes were seen for a given temperature and field.[27,34] In each case the difference arose when detecting the spin of the electrons through the trion transition vs. through the exciton transition: a factor of 6 difference in lifetime for Ref. 27 and a factor of 2 for Ref. 34. Each group attributed the difference in lifetimes to a difference in localization of the subset of electrons being probed: localized electrons in the case of the trion transition and quasi-free electrons in the case of the exciton transition. The trion and exciton transitions are clearly resolvable in II-VI quantum wells, but in our GaAs quantum well the light hole trion transition and heavy hole exciton were likely both contained in the "light hole trion region" marked on Fig. 2 (which is where we probed). Nevertheless, we may have seen this effect in our $T_1$ measurements as well. Fig. 7 displays the results of a fine-scale wavelength adjustment: measuring the spin lifetimes as we varied the probe laser across the light hole trion peak. The trion peak from the non-time-resolved wavelength-dependent Kerr rotation signal is shown dashed. As the wavelength was tuned from one side of the peak to the other, there was an abrupt shift in spin lifetime. It seems likely that this shift in lifetime is a result of probing different subsets of electrons (e.g. localized vs. quasi-free) electrons, just as was seen in the II-VI quantum well experiments.

## VII. CONCLUSION

In conclusion, we have measured $T_1$ spin flip times in a GaAs quantum well by tuning a probe laser to be resonant with the optical transition of the well in a longitudinal (Faraday) geometry. The well had extremely long spin lifetimes, exceeding one microsecond for 1.5 K and 5.5 T. This quite likely indicates long $T_2$ lifetimes as well. Lifetimes increase with field, and decrease with temperature. Nuclear polarization effects were significant at the highest fields at 1.5 K, but could largely be removed with nuclear magnetic resonance. Different lifetimes were observed with small changes in wavelength for one set of experimental conditions, likely indicating responses from two differently-localized subsets of electrons.